\documentclass[letterpaper]{jpconf}
\usepackage{graphicx}
\usepackage{subfigure}

\def \Eslash {E \kern-.5em\slash }
\def \pslash {p \kern-.5em\slash }
\def \kslash {k \kern-.5em\slash }

\newcommand{\GeV}{\ensuremath{\mathrm{Ge\kern -0.1em V}}}
\newcommand{\TeVcc}{\ensuremath{\,\mathrm{Te\kern -0.1em V\!/c}^2}}
\newcommand{\GeVcc}{\ensuremath{\,\mathrm{Ge\kern -0.1em V\!/c}^2}}
\newcommand{\MeVcc}{\ensuremath{\,\mathrm{Me\kern -0.1em V\!/c}^2}}
\newcommand{\GeVc}{\ensuremath{\mathrm{Ge\kern -0.1em V}\!/c}}

\newcommand{\ifb}{\ensuremath{\mathrm{fb^{-1}}}}

\newcommand{\Prot}{\ensuremath{\mathrm{p}}}

\newcommand{\PP}{\Prot\Prot}

\newcommand{\ttbar}{\ensuremath{\mathrm{t}\bar{\mathrm{t}}}}

\begin{document}
\title{Higgs search in H$\rightarrow$ZZ/WW decay channels with the CMS detector}

\author{C.~Charlot, for the CMS collaboration}

\address{Laboratoire Leprince-Ringuet, Ecole Polytechnique and IN2P3/CNRS,  Palaiseau, France}

\begin{abstract}
A prospective analysis for the search of the Standard Model Higgs
boson decaying in vector boson pairs is presented with the CMS experiment
in the context of the initial luminosity at the CERN LHC \PP\;collider.
Monte Carlo data corresponding to an integrated luminosity of
up to 1~\ifb\;are analysed and the expected significance for
a Standard Model-like  Higgs boson in these channels is established.
\end{abstract}

\section{Introduction}

The Standard Model (SM) of electroweak and strong interactions predicts the existence
of a single physical Higgs boson, the quantum of the scalar field responsible for electroweak
symmetry breaking. 
Direct searches for the SM Higgs particle at the LEP $\rm e^+e^-$ collider
have lead to a lower mass bound of $m_{\rm H} > 114.4 \GeVcc$ (95\% CL)~\cite{LEPEWWG}.
Ongoing direct searches at the TeVatron $\rm  p\bar{p}$ collider by the D0 and CDF
experiments set constraints on the production cross-section for a SM-like Higgs boson
in a mass range extending up to about $200 \GeVcc$~\cite{CDFD0Higgs, CDFD0New}.
A consistency fit including all the measured electroweak observables
%sensitive to the existence of a Higgs boson
favours the mass range $m_{\rm H} < 182 \GeVcc$ (95\% CL)~\cite{EPS07Higgs}. 
The inclusive production of SM Higgs bosons followed by the decay into di-bosons and subsequently into leptons, $
{\rm H} \rightarrow {\rm Z} {\rm Z}^{(*)} \rightarrow l
l l' l'$ and ${\rm H} \rightarrow {\rm W} {\rm W}^{(*)} \rightarrow l
\nu_{l} l' \nu_{l'}$ with $l, l' = e$ or $\mu$, are expected to
be early discovery channels at the CERN LHC $pp$ collider over a
wide range of possible $m_{\rm H}$ values. 
We present here the analysis strategies for the the Higgs search in di-boson decay channels and leptonic modes in the context of
an initial luminosity of 1~\ifb. 
Signal and background datasets obtained with a detailed Monte Carlo
simulation of the detector response, including the limited inter-calibration and alignment precision expected at startup, are treated using a complete reconstruction chain.
Emphasis is put on the reduction
of distinguishable background rates and on methods allowing for a
data-driven derivation of experimental and background systematic
uncertainties. 

\section{The CMS detector}

A general description of the CMS detector can be found elsewhere~\cite{CMSDetector}. This analysis relies mostly on the
tracker and the electromagnetic calorimeter
(ECAL), both immersed in a 4\,T magnetic field
parallel to the $z$ axis, and on
the muon spectrometer hosted in the iron magnet
return yoke.
The CMS tracker is a cylindrical detector equiped with silicon pixel detectors 
for the innermost part and silicon strip detectors for the outer layers. 
The tracker acceptance for a minimum of 5 collected hits extends up to 
pseudorapidities $\eta$ of about $\vert \eta \vert < 2.4$. 
The CMS ECAL is made of quasi-projective PbWO$_4$ crystals with a granularity of approximately $\Delta \eta \times \Delta \phi = 0.0175 \times 0.0175$ in the barrel part ($\vert \eta \vert \leq 1.48$) and of approximately $\Delta x \times \Delta y = 1.3R_M  \times 1.3R_M$ in the endcaps parts. The endcaps are equipped with a preshower device that cover the region
$ 1.6 < \vert \eta \vert < 2.6 $. 
The electron reconstruction efficiency varies from 85\% to 95\% depending on $p_T$ and $\eta$.
The muon spectrometer consists of Drift Tubes (DTs),
Cathode Strip Chambers (CSCs),  and Resistive Plate Chambers (RPCs), 
which cover the angular region $\vert \eta\vert < 2.4$. The muon reconstruction efficiency varies between 95\% and 99\%.

\section{The WW$^{(*)}\rightarrow l\nu_{l} l' \nu_{l'}$ analysis}

The analysis in this channel is performed using either a simple cut based approach and a more involved multi-variate analysis based on a neural network (NN). 
%Events are caracterized by the presence of two isolated high ${p_\mathrm{T}}$ leptons plus a significant missing transverse energy from the undetected neutrinos, together with a small hadronic activity in the central part of the detector.
Events passing leptonic trigger paths with two identified and isolated high-${p_\mathrm{T}}$ leptons ($e$ or $\mu$) are
selected. Standard lepton reconstruction techniques are used, and
the identification of electrons is relatively tight to reduce
the contamination from $W$+jets processes.
The charged isolated leptons
identified in this way are combined into all possible pairs requiring leptons of opposite charge  and within
$|\eta|\leq 2.5$ and ${\ensuremath{p_T}} \geq 10\;\mathrm{GeV}$ or 
at least one lepton with ${\ensuremath{p_T}} \geq 20\;\mathrm{GeV}$.
If none or more than one such pair is found 
the event is rejected so to suppress WZ and ZZ backgrounds.
A minimum missing transverse energy of 30 GeV is required in accordance with the
presence of two neutrinos in the final state. 
An event containing any jet with ${p_\mathrm{T}}>15$~GeV and $|\eta| < 2.5$ is rejected.
This cut removes the bulk of the \ttbar~background.
Finally, $m_{\ell\ell}\;>\;12\;GeV$ is required to select events with leptons coming
from fully leptonic W~pair decays.
Additional variables are used in the final kinematical selection both 
in the cut based and in the NN analysis: the angle $\Delta \Phi_{\ell\ell}$ between the two leptons in the transverse plane;
the invariant mass of the lepton pair ($e^+e^-$ and $\mu^+\mu^-$ final states);
the missing transverse energy
and the transverse momenta of the harder and the
softer lepton.
In the NN analysis, following additional variables are used: the separation angle $\Delta \eta_{\ell\ell}$ between the leptons in $\eta$, the transverse mass of both lepton-missing transverse energy pairs, the $\eta$ angle of both leptons, the angle in the transverse plane between the missing transverse energy and the closest lepton and the flavour of the di-lepton final state.
Systematic uncertainties play an important role in this analysis
where no strong mass peak is expected due to the presence of two neutrinos in the final state.
Experimental systematic uncertainties coming from the luminosity measurement, lepton identification and efficiencies, missing transverse energy resolution, jet efficiency and energy scale have been taken into account. 
The normalization of the two main backgrounds, \ttbar~and ${\rm W}^+{\rm W}^-$  has
been addressed using various data driven methods. 
The overall relative error depends on the final state and on the Higgs mass.
It is estimated as about 11\% for the signal and 21\% for the background.
Fig.~\ref{fig:tmva_outputs} (left) shows the neural network outputs for the signal and the backgrounds for $m_{\rm H}=170~\GeVcc$. 
The distributions are representative of other mass regions. 
There is a clear shape difference between signal and background events, although there is no region completely free of background. The vertical line indicate the cut value used.

\section{The ZZ$^{(*)}\rightarrow lll'l'$ analysis}

This channel is caracterized by the presence of two pairs
of isolated primary electrons or muons, with one pair generally
resulting from the decay of a Z boson on its mass shell. 
After the High Level Trigger, the event rates in the
lepton paths are still dominated by "fake" leptons coming
predominantly from QCD processes. 
To reduce the contribution of QCD multijets and
Z/W+jet(s)  at a level comparable or below the contribution
of the three main backgrounds, $t\bar{t}$, ${\rm Z}b\bar{b}$ and
${\rm Z}{\rm Z}^{(*)}$, a set of  pre-selection cuts is applied.
Events are kept only if
at least one combination of two matching pairs is found with an
invariant mass greater than $100 \GeVcc$.
%, thus restricting the phase
%space for the search of the SM Higgs boson to non-excluded
%$m_{\rm{H}}$ range. 
To further suppress  the Z+jet(s) contamination,
a loose track-based isolation is applied. 
The major reducible backgrounds then remaining 
%after pre-selection 
are
Z+jet(s), $t\bar{t} \rightarrow {\rm W}^+ b {\rm W}^- \bar{b}$ and
${\rm Z}b\bar{b}$ with fake leptons from jets or
semi-leptonic decays of bottom mesons. 
%These leptons are
%likely to be accompanied by hadronic products from the fragmentation
%and decay processes initiated in the light-quarks or b-quark jets.
%Moreover, the leptons from the $b$-jets in $t\bar{t}$ and ${\rm
%Z}b\bar{b}$ background events are likely to have a large impact
%parameter with respect to the primary vertex because of the long
%lifetime of the $b$-hadron. Thus, a 
The final selection incorporates tighter
lepton isolation complemented by three-dimensional impact  parameter measurements.
In order to best preserve the signal
detection efficiency while acting on low $p_T^{\ell}$ lepton
candidates to suppress the ${\rm Z}b\bar{b}$ background, the
isolation criteria for the leptons from the pair of lowest
$m_{\ell^{+}\ell^{-}} $ is made $p_T^{\ell}$ dependent. 
The $m_{2\ell}$
masses observables are finally exploited with very loose
cuts to preserve the simplicity of an $m_{\rm H}$ independent 
selection for the initial luminosity: it is required a reconstructed "Z" with $50 <
m_{\rm Z} < 100 \GeVcc$ and a "Z*" with $20 < m_{\rm Z} < 100
\GeVcc$.
Contrary to the WW$^{(*)}$ channel, the ZZ$^{(*)}$ decay mode allows in presence of signal for a narrow peak in the four lepton invariant mass distribution
%, which is therefore the most discriminating observable in this channel.
as shown in Fig.~\ref{fig:tmva_outputs} (right). 
The Z+jet(s) and $t\bar{t}$ backgrounds are completely eliminated.
The ${\rm Z}b\bar{b}$ background is considerably reduced and
only survives towards low masses.
%, with an event rate far below that
%of the ${\rm Z}{\rm Z}^{(*)}$ continuum. 
\begin{figure}[!ht]
\begin{center}
\includegraphics[viewport=100 10 670 530,width=0.4\textwidth,height=.23\textheight]{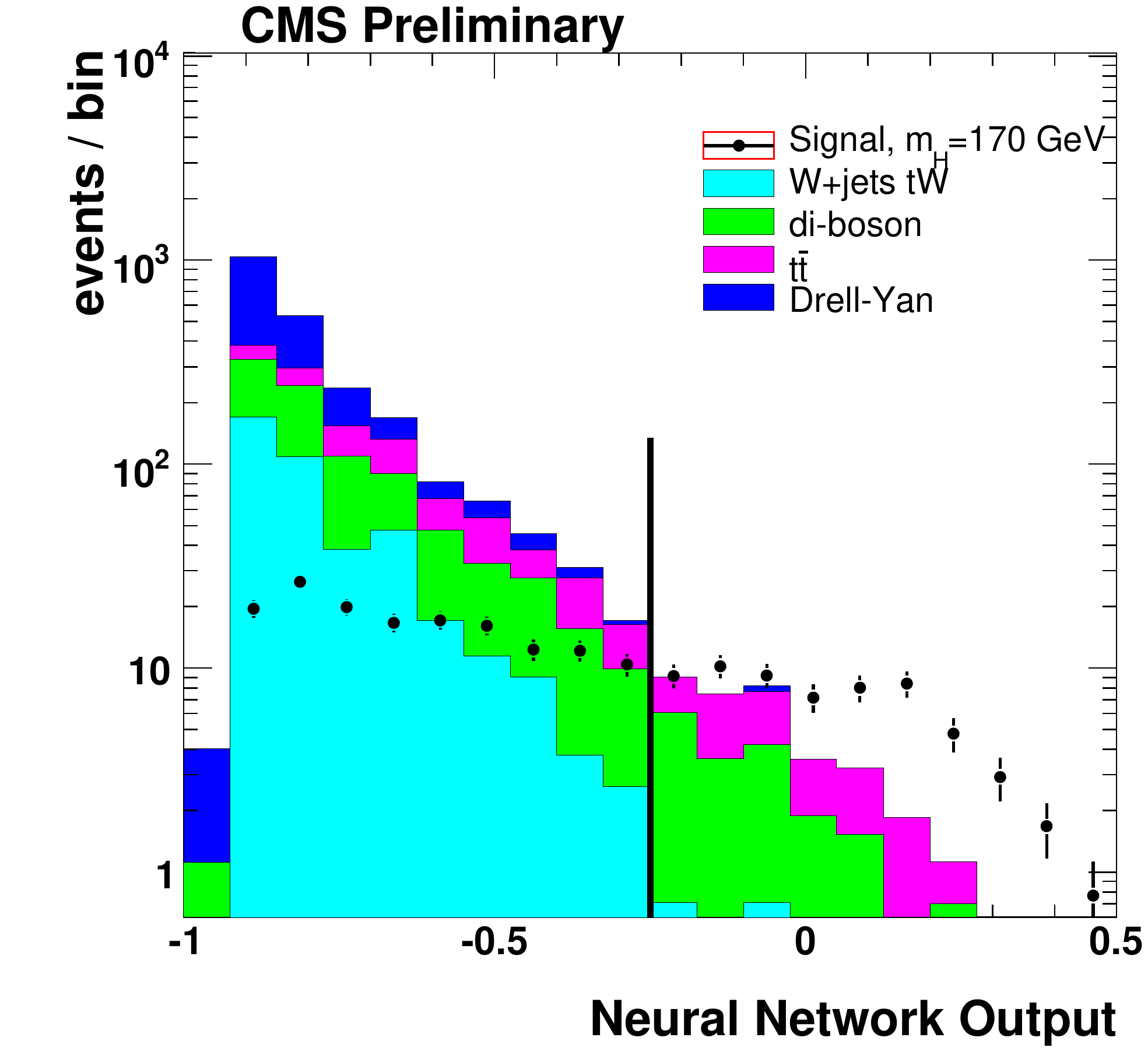}
\includegraphics[viewport=0 10 500 480,width=0.36\textwidth,height=.2\textheight]{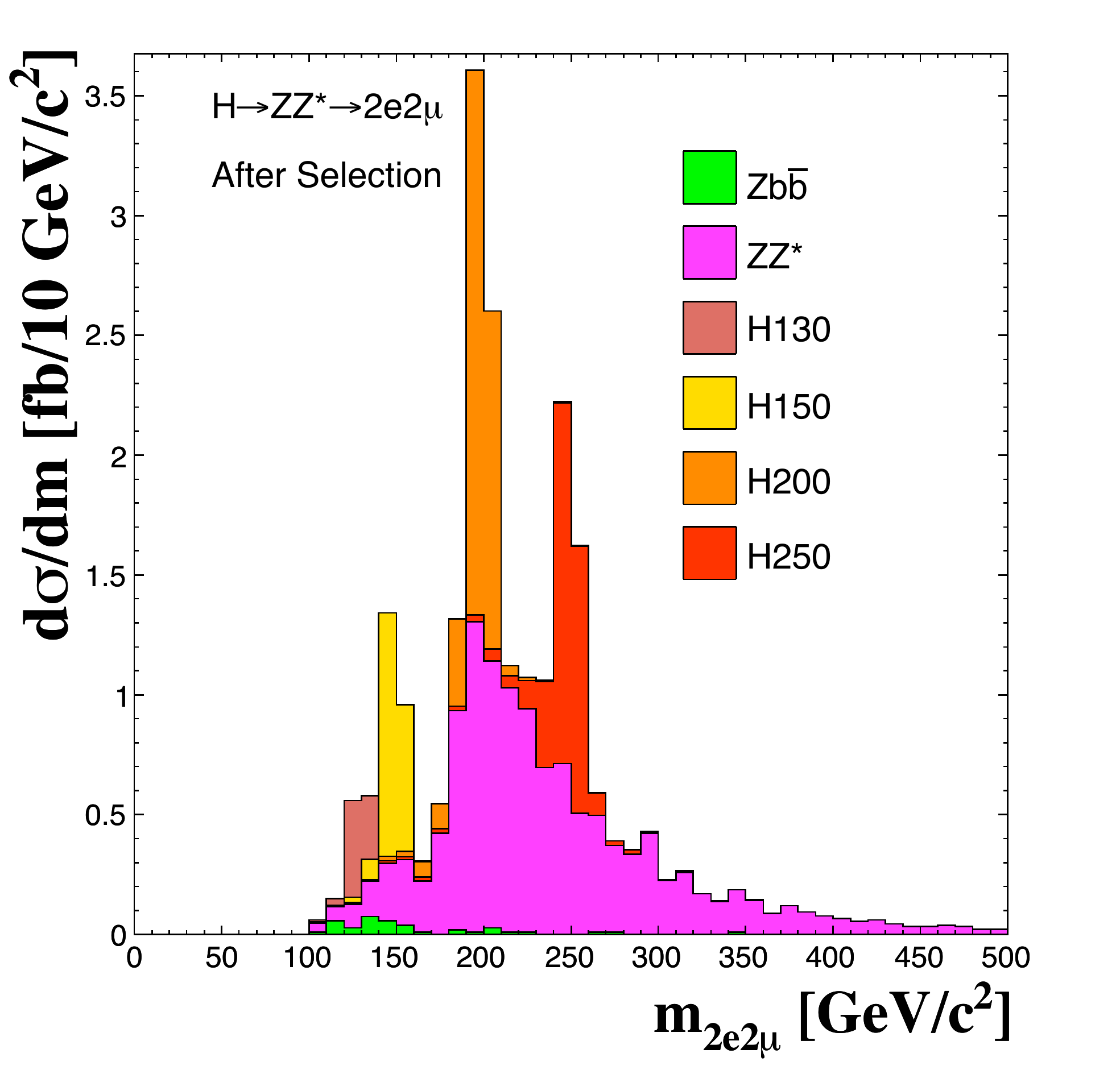}
\caption{left: NN outputs for signal (blue squares) and background (red
circles) in the WW$^{(*)}$ channel for
\mbox{$m_{\rm H}=170~\GeV$}; right: four lepton invariant mass after baseline selection in the ZZ$^{(*)} \rightarrow 2e2\mu$ channel. Results are for an integrated luminosity of 1~\ifb. }
\label{fig:tmva_outputs}
\end{center}
\end{figure}
\section{Results}
The same likelihood ratio technique as used at LEP and the
Tevatron~\cite{likrat} is used here to evaluate the significance of the experiment to the presence 
of a Higgs boson signal. 
The expected significance of an event excess under the  assumption
of the presence of a Higgs boson is shown in 
Fig.~\ref{fig:RejectionLimits} (left) for the WW$^{(*)}$ channel and
for the NN 
analysis. Systematic uncertainties for the signal and background
are taken into account. A SM Higgs boson can be found at 5$\sigma$ in this channel around 
\mbox{$m_{\rm H}=160~\GeV$}.
In the case of the ZZ$^{(*)}$ channel, the uncertainty on the
observations is dominated by statistical
fluctuations for the considered integrated luminosity. 
It is found unlikely that an
integrated luminosity of 1~\ifb will yield an observation of a
mass peak with a significance well above 2$\sigma$.
\begin{figure}[tbh]
\begin{center}
\includegraphics[viewport=30 40 600 550,width=0.45\textwidth,height=.243\textheight]{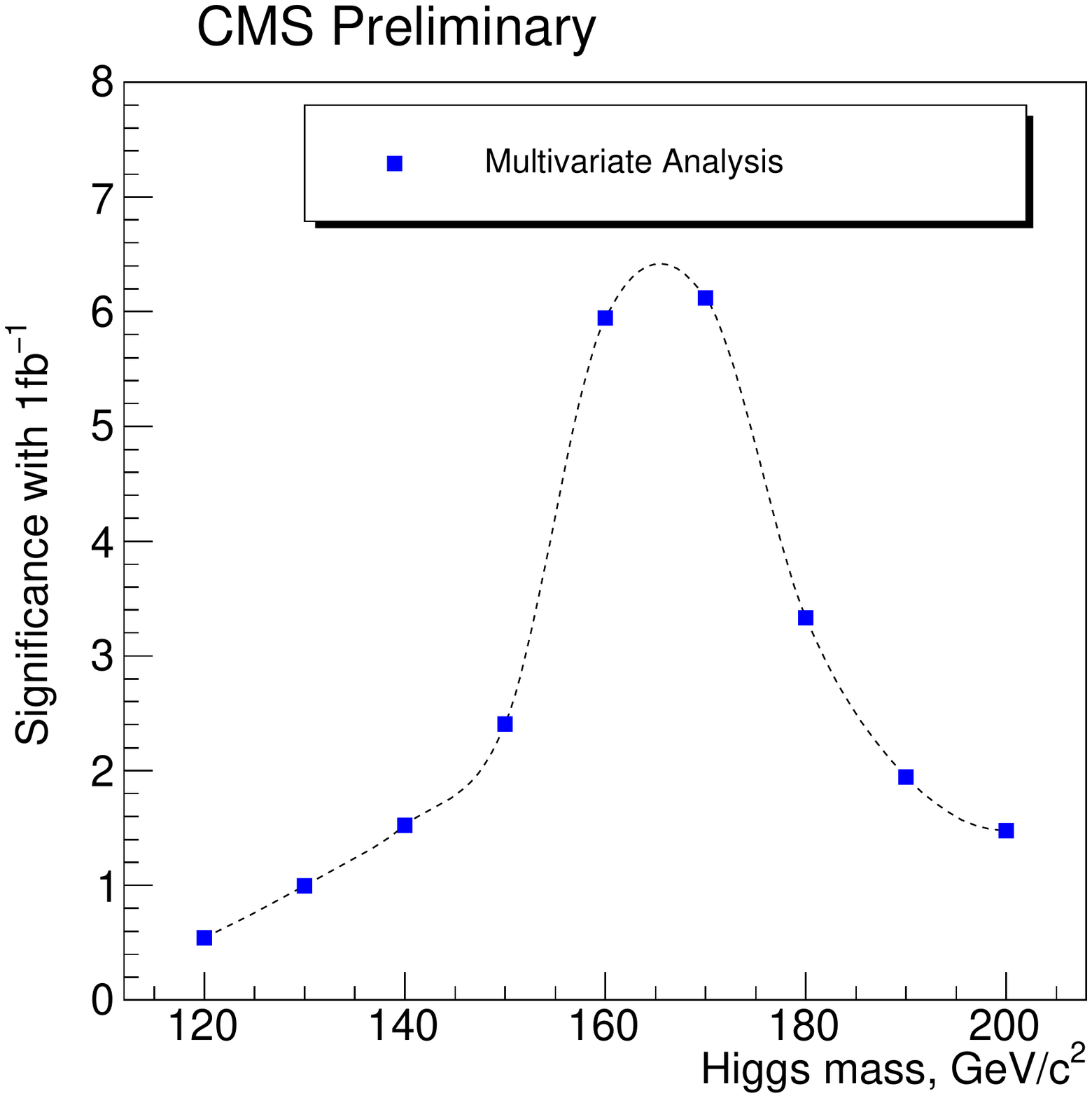}
\includegraphics[viewport=30 60 700 555,width=0.45\textwidth,height=.243\textheight]{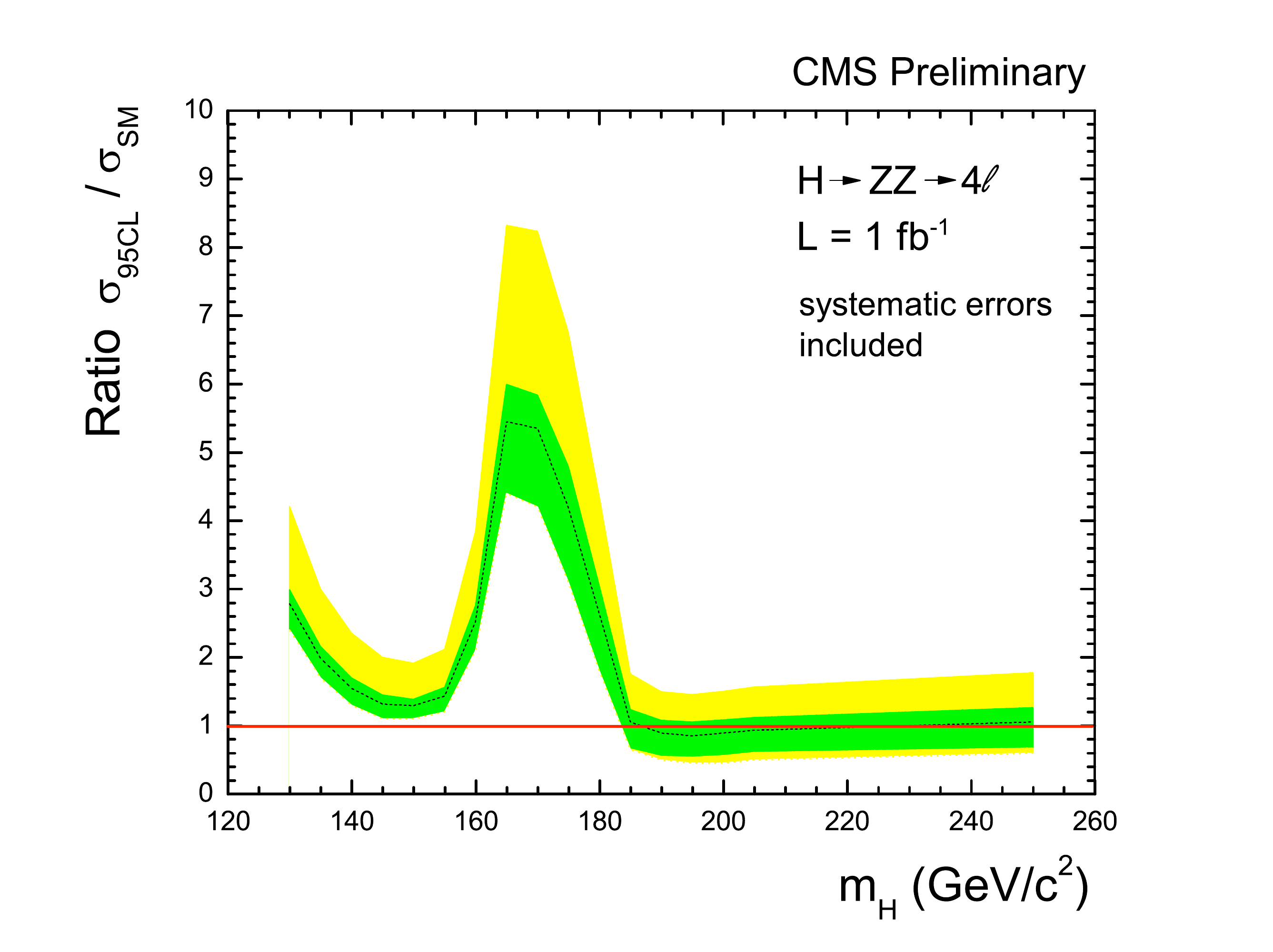}
\end{center}
 \caption{left: expected significance as obtained from the NN analysis for the WW$^{(*)}$ channel; right: expected excluded cross-sections for a SM-like Higgs boson normalized to the SM Higgs boson cross sections.
               The green (yellow) bands shows the
               68\% (95\%) coverage range. Results are for an integrated
               luminosity of 1~\ifb.}
               \label{fig:RejectionLimits}
\end{figure}
In absence of a significant deviation from the SM expectations, an
upper limit on the cross-section for the production of a SM-like
Higgs boson can be derived.
% using a Bayesian approach. 
The results are presented in
Fig.~\ref{fig:RejectionLimits} (right) for various $m_{\rm H}$
hypothesis and expressed in terms of the ratio of excluded over
Standard Model cross sections $R_{95\%C.L.} = \sigma_{95\%C.L.} /
\sigma_{SM}$. Systematic errors on the signal and background are taken into account, with an assumption
of their 100\% correlation. One can see that there is a fair chance of excluding
the SM-like Higgs at 95\%C.L. for the mass range $185-250\GeVcc$.
From the event counts for signal and background, together with associated systematic errors, it is possible to combine results of expected sensitivity in all di-boson decay channels. 
The combination of these channels is done using two different approaches, a bayesian calculation and a method based on confidence levels, with different assumptions on the correlation between systematic errors. As a result, combining the di-boson decay modes in the leptonic final states, a SM-like Higgs boson can be excluded at the LHC with 1~\ifb~for all masses above $\sim$ $140\GeVcc$.
Finally, a preliminary evaluation of the combined sensitivity in the case of a 10 TeV initial energy of the LHC has been performed. The dominant effect is the change in cross-sections due to partonic luminosities. 
%Reducing the c.m. energy leads to more loss in the signal than in the background yields due to the gluon-gluon luminosity decreasing faster than the quark-anti quark luminosity. 
Using a simple rescaling of the signal and background yields leads to  a loss of about a factor 1.5 in sensitivity going from 14 TeV to 10 TeV. Detailed analyses at 10 TeV involving a complete reevaluation of systematics are ongoing.

\section{Conclusions}

The CMS experiment is actively preparing for the Higgs search with complete analyses based on detailed Monte carlo simulations and full event reconstruction. The di-boson decay modes with their leptonic final states represent the main discovery channels for integrated luminosities  of around 1~\ifb. A SM Higgs can be found at 5$\sigma$ in the WW$^{(*)}$ channel around
\mbox{$m_{\rm H}=160~\GeV$} for an integrated luminosity of 1~\ifb. 
Due to its lower  $\sigma \times$BR, the ZZ$^{(*)}$ channel doesn't allow for a discovery with such integrated luminosities 
and 95\% confidence level exclusions limits can be set in absence of deviation from SM expectation.
Combining both channels, it is found that with an integrated luminosity of 1~\ifb~a SM-like Higgs boson can be excluded at the LHC for all masses above $\sim$140 $\GeVcc$. In the case of a 10TeV initial energy of the LHC machine, the sensitivity would decrease by a factor of about 1.5.

%\vspace{0.5cm}
%%\clearpage
%We thank the technical and administrative staff at CERN and other 
%CMS Institutes, and acknowledge support from:
%FMSR (Austria); 
%FNRS and FWO (Belgium); 
%CNPq, CAPES, FAPERJ and FAPESP (Brazil); 
%MES (Bulgaria); 
%CERN; 
%CAS, MST and NSFC (China); 
%MST (Croatia); 
%RPF (Cyprus); 
%Academy of Sciences and NICPB (Estonia); 
%Academy of Finland, ME and HIP (Finland); 
%CEA and CNRS/IN2P3 (France); 
%%BMBF and DESY (Germany); 
%BMBF, DFG and HGF (Germany); 
%GSRT and Leventis Foundation (Greece); 
%OTKA and NKTH (Hungary); 
%DAE and DST (India); 
%IPM (Iran); 
%SFI (Ireland); 
%INFN (Italy); 
%KICOS (Korea); 
%CINVESTAV, CONACYT, SEP and UASLP-FAI (Mexico); 
%PAEC (Pakistan); 
%SCSR (Poland); 
%FCT (Portugal); 
%JINR (Armenia, Belarus, Georgia, Ukraine, Uzbekistan);
%MST and MAE (Russia);
%MSD (Serbia);
%MCINN and CPAN (Spain); 
%Swiss Funding Agencies (Switzerland);
%NSC (Taipei); 
%TUBITAK and TAEK (Turkey); 
%STFC (United Kingdom); 
%DOE and NSF (USA).

\bibliographystyle{aipproc}   % if natbib is available

\end{document}